\documentclass[aps,prb,twocolumn,showpacs,showkeys]{revtex4}

\usepackage{graphicx}

\begin{document}

\newcommand{\psides}{\{211\}}

\title{Observation of the vertex-rounding transition for a crystal in equilibrium:
oxygen-covered tungsten}

\author{Andrzej Szczepkowicz}
\author{Robert Bryl}

\affiliation{Institute of Experimental Physics, University of Wroc{\l}aw,
Plac Maksa Borna 9, 50-204 Wroc{\l}aw, Poland}

\date{\today}

\begin{abstract}
Equilibrium crystal shape of oxygen-covered tungsten is followed as a
function of temperature using field ion microscopy. 
In the vicinity of the (111) region, at the temperature $970\pm70$~K, the system undergoes 
a phase transition from a polyhedral form (sharp edges and sharp vertex)
to a rounded form (sharp edges, rounded vertex).
\end{abstract}

\pacs%
{%
 68.35.Md 
 68.60.Dv 
 68.37.Vj 
}


\maketitle

\section{Introduction\label{introduction}}

\begin{figure}
\includegraphics{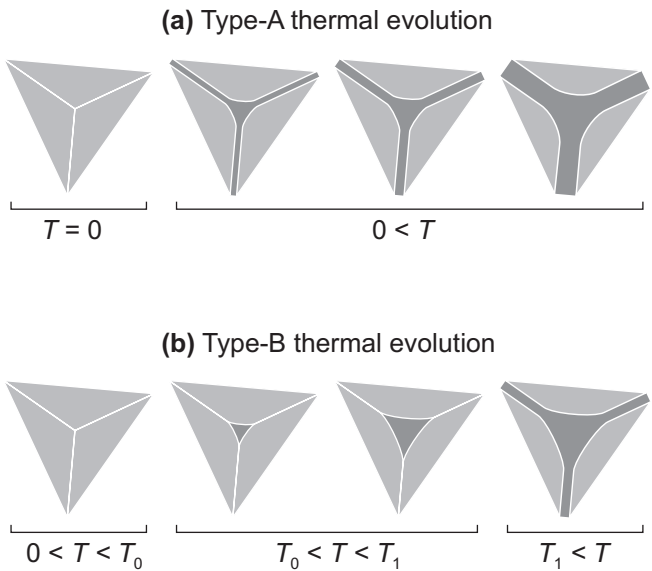}%
\caption%
{%
 \label{Fig-wortis}%
 Thermal evolution of a vertex of equilibrium crystals \cite{RottmanWortis1984,Wortis1988}.
 Dark shading denotes the rough, rounded region.
 In (b), $T_0$ denotes the vertex-rounding temperature, and $T_1$ denotes
 the edge-rounding temperature.
}
\end{figure}

The shape of a crystal in equilibrium depends on the temperature.
Theoretical considerations suggest that at 0~K equilibrium crystal 
shapes (ECS) are polyhedral \cite{Wortis1988,FrenkenStoltze1999}. 
On the other hand, 
both theory and experiment show that the ECS are almost spherical
near the melting point. The detailed evolution of ECS with temperature
depends on the particular system studied. According to the
three-dimensional Ising model with nearest-neighbor and 
next-nearest-neighbour interaction
\cite{Wortis1988,RottmanWortis1984,JayaprakashSaam1984,TouzaniWortis1987,ShiWortis1988}
there appear to be two basic evolution patterns
depicted in Fig.~\ref{Fig-wortis}. 
In type-A evolution \cite{RottmanWortis1984,Wortis1988}, the crystal is polyhedral
only at 0~K. At arbitrarily low nonzero temperature 
facets are separated by rough, rounded regions; 
no sharp edges or vertices are present on the surface.
In type-B evolution (obtained for repulsive next-nearest-neighbor interaction) 
\cite{RottmanWortis1984,Wortis1988}, 
there exist two characteristic temperatures,
the vertex-rounding temperature $T_0$ and the edge-rounding
temperature $T_1$. The crystal remains strictly polyhedral up to 
$T_0$. Between $T_0$ and $T_1$ the vertices are smoothly rounded, but
the edges remain sharp. Above $T_1$ the sharp edges disappear.
For high temperatures, both models predict the disappearance of flat facets
at the roughening temperature (facet-rounding temperature).

In experiment, the two evolution patterns are difficult to distinguish,
because at low temperature it is often impossible to achieve equilibrium
due to material transport limitations,
while at high temperature the two patterns are similar.
Nevertheless, experiments suggest that $^4$He crystals follow type-A
evolution \cite{TouzaniWortis1987}, 
while NaCl crystals follow type-B evolution
(although no images of ECS between $T_0$ and 
$T_1$ have been obtained) \cite{ShiWortis1988}.

In this paper we describe how the equilibrium shape of
oxygen-covered tungsten (O/W) changes with temperature. We show
that the thermal evolution in the vicinity of the [111]-oriented 
vertex is of type-B,
and estimate the vertex-rounding temperature $T_0$.

\section{Experimental}

\label{experimental}

In this experiment the equilibrium crystal was approximated by a
[111]-oriented, needle-shaped tungsten crystal (field emitter geometry),
with the cone half-angle not exceeding 10$^{\circ}$.
The apex of the needle was approximately 
hemispherical, with the average radius of curvature $250$~nm
(crystals of smaller radii do not survive thermal cleaning, while
larger crystals are difficult to equilibrate).
With such crystal geometry, for orientations within $25^{\circ}$
of the central direction (in our case: [111]),
the average radius of curvature is constant up to $\pm3$~\% \cite{NicholsMullins1965}.
Therefore, in the vicinity of the (111) pole, the needle crystal is a good
approximation of the true equilibrium shape.
In this study we concentrate only on the vicinity of the (111) pole.

The experiment was carried out in a field ion microscope 
\cite{FIM-Mueller-Tsong,FIM-Oxford}, with the base pressure of $3\cdot10^{-10}$~Torr.
During the thermal shaping of the crystal, the average pressure was higher: 
$1\cdot10^{-9}$~Torr, due to the residual presence of the gases 
used for imaging (helium, neon, krypton) and for adsorption (oxygen).

To observe the equilibrium shape at a temperature $T$, the following procedure was applied: 
(1) The crystal was cleaned thermally \cite{SzczepkowiczCiszewski2002}.
(2) The crystal was cooled to 80~K and its surface was covered by oxygen ($1.4\pm0.3$~L).
(3) The crystal was pre-annealed at 1500~K for 80~s to achieve a globally faceted 
shape \cite{SzczepkowiczBryl2004}.
(4) For technical reasons, the crystal was for a short period ($\sim$1 min) 
held at low temperature (near 80~K).
(5) The crystal was equilibrated at the temperature of interest $T$ for 80~s.
(6) The $T$ equilibrium configuration was frozen by rapid cooling
of the crystal to 80~K (the estimated quenching rate is 400 K/s at 1000~K).
(7) The crystal surface around the (111) crystal pole was observed 
by field ion microscopy.

To observe the ECS at a different temperature, the whole procedure was repeated
(including crystal cleaning), to avoid possible contamination of the surface
by residual gases.

\section{Results\label{results}}

\begin{figure}
\includegraphics{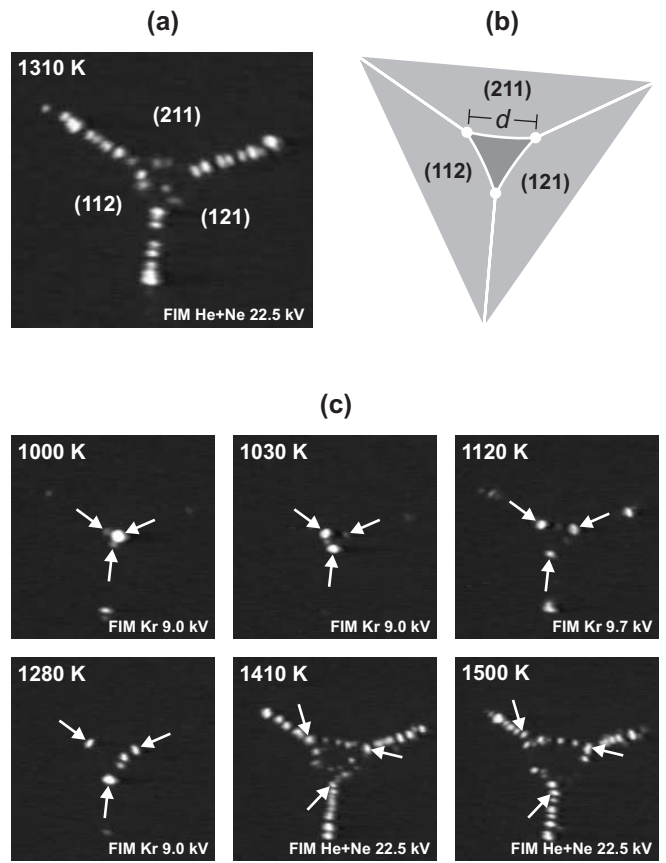}%
\caption%
{%
 \label{Fig-fim}%
 Thermal evolution of an oxygen-covered tungsten crystal.
 (a)~Microscopic image of the vicinity of the (111) pole.
 The \psides\ facets form a pyramid pointing in the [111] direction.
 (b)~A model corresponding to~(a). The degree of vertex rounding 
 can be characterized by the distance $d$ between the edge end points.
 (c)~The distance between the edge end points (indicated by arrows) 
 increases with temperature.
}
\end{figure}

Figure \ref{Fig-fim}~(a) shows an example microscopic image
of the crystal equilibrated at $T=1310$~K. 
Only the area around the (111) pole is shown.
The surface configuration seen in the image is a good approximation of the ECS,
with possible minor differences due to nonzero time of quenching. The edges connecting the
\psides\ facets are sharp on the atomic scale, but the (211)-(121)-(112) pyramid is incomplete
-- the vertex is missing. In Fig. \ref{Fig-fim}~(b) we propose a corresponding model of
the surface, with a rough, rounded vertex region. Figure \ref{Fig-fim}~(c) demonstrates that 
the area of the rough region increases
with temperature, in accord with the theory of ECS \cite{Wortis1988}.

\begin{figure}
\includegraphics{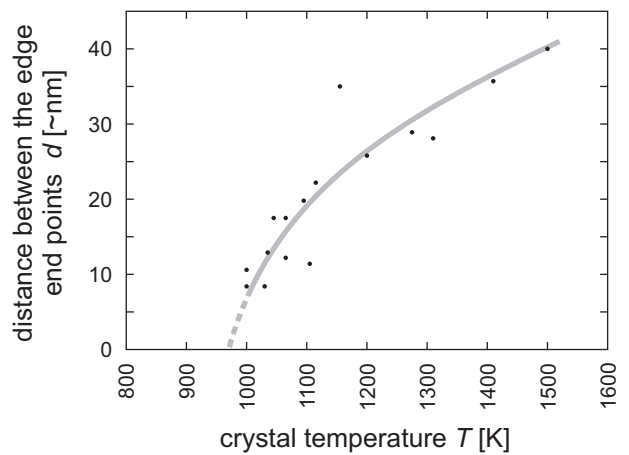}%
\caption%
{%
 \label{Fig-plot}%
 The size of the rough region at the (111) crystal pole,
 characterized by the distance $d$ between the edge end points,
 increases with temperature.
}
\end{figure}

In order to demonstrate better the changes of the rough area with temperature,
we characterize the degree of vertex rounding by measuring the distance $d$ between
the edge end points (Fig.~\ref{Fig-fim}~(b)). This distance is plotted in
Fig.~\ref{Fig-plot} against the temperature. 
In measuring $d$, we have neglected
local variations in magnification of the microscope 
\cite{FIM-Oxford, SzczepkowiczCiszewski2002}; for this reason 
the nanometer scale in Fig.~\ref{Fig-plot} is only approximate,
and the lengths are probably overestimated for low values of $d$. By slight extrapolation
of the plotted data, we obtain the vertex-rounding temperature $T_0=970\pm70$~K.

\section{Discussion\label{discussion}}

We have studied the thermal evolution of an 
\emph{adsorbate-covered} metal crystal, a system more complex than a 
-- already difficult to describe theoretically -- \emph{pure} metal crystal.
Is the presence of an adsorbate essential for the 
vertex-rounding transition to occur on real crystals?
The answer is most probably no. However, oxygen adsorption on tungsten greatly increases
the anisotropy of the surface free energy in the vicinity of the (111) pole.
In effect, while between 1000 K and 1200 K the \psides\ facets of 
pure tungsten are separated by rounded regions
\cite{Bassett1965,SzczepkowiczCiszewski2002}, the
oxygen-covered tungsten develops sharp edges
between the \psides\ facets (Fig.~\ref{Fig-fim}), and most probably 
a sharp (111) vertex below $\sim$970 K (Fig.~\ref{Fig-plot}).
It is possible that \emph{pure} tungsten also undergoes a vertex-rounding
transition of similar type, but at much lower temperature,
where equilibrium cannot be achieved by surface self-diffusion.

The oxygen coverage, corresponding to 1.4 L exposure applied in this study,
does not exceed $5\cdot10^{14}$ molecules/cm$^2$.
At such coverage, desorption of oxygen at temperatures not exceeding 1500~K 
is neglegible \cite{KingMadeyYates1971}, so the adsorbate film during
the thermal evolution of the crystal is stable.

Oxygen adsorption on flat W(111) crystal face has been described
in the literature
\cite{Taylor1964,TracyBlakely1968,Madey1990,Madey1993a}. 
At temperatures above 800 K
the O/W(111) system is unstable and undergoes massive
hill-and-valley faceting, exposing \psides\ facets.
According to the theory of ECS \cite{Herring1951,Wortis1988}, 
this means that, at least at 800 K, the (111) facet
is absent from the equilibrium form of O/W, which is consistent
with the results presented here.

At the transition temperature, the (111) vertex is surrounded by
three \psides\ facets. The adsorption of oxygen on 
W(211) has been described in the literature 
\cite{ChangGermer1967,TracyBlakely1969,HopkinsWatts1974,ErtlPlancher1975,
WangLu1983,WangPimbleyLu1985,WangLu1985,Bu1989}. 
Oxygen exposure of 1.4 L causes the formation of 0.5--1.5 geometric monolayers 
of oxygen atoms on W(211).
At low temperatures, the adsorbate forms ordered phases:
$p(2\times1)$, $p(1\times1)$ and $p(1\times2)$, depending
on the coverage. However, above 900 K the adsorbate
is disordered at all coverages -- it forms a two-dimensional 
lattice gas \cite{WangLu1983,WangLu1985}.

In Section \ref{introduction}, we have put our results in the 
context of models of thermal evolution of \emph{pure} crystals.
To the best of our knowledge, in the literature so far there are 
no reports of models of thermal evolution of ECS with adsorbate
(although zero-temperature models of ECS with adsorbate have been described
\cite{Shi1987,Shi1988}).
It should also be noted that such small crystals as the one
studied here possibly can display finite-size effects, in contrast
to the often studied ,,thermodynamic limit'' case \cite{Wortis1988}.
Recently, a simple Monte Carlo solid-on-solid BCC model has been described
\cite{Oleksy2004}, which is in accord with certain features of the experimentally 
observed adsorbate-induced faceting. The extension of this model to
curved surfaces is under development \cite{Oleksy2004s}.

To the best of our knowledge, the thermal evolution of the
ECS vertex described in Section~\ref{results}
has not yet been observed for pure or adsorbate-covered metal crystals.
The edges of ECS of pure metals always appear rounded in the microscopic images
\cite{Bonzel2003}. It seems that such crystals either exhibit type-A evolution,
or the temperatures studied are higher than the edge-rounding temperature.
On the other hand, annealed adsorbate-covered crystals often develop sharp
edges between the facets and sharp vertices 
\cite{Drechsler1983, NienMadey1997, Pelhos1999}, 
but the temperature-dependent
vertex rounding has not been reported. The sharpness of the vertices
observed in some adsorbate-induced faceting experiments may be due to low annealing 
temperatures (below the vertex-rounding temperature), or due to low quenching rates
(non equilibrium effect: the vertex is rebuilt during cooling).

In our study, 
the average radius of curvature of the crystal (see Section \ref{experimental}) 
is an order of magnitude smaller than in a typical ECS experiment \cite{Bonzel2003}. 
Because of this small size, thermal equilibration of the crystal is possible at
relatively low temperatures. We have found that 80 seconds of annealing
is sufficient to obtain equilibrium, at least around the (111) pole,
already at 1400~K \cite{SzczepkowiczBryl2004}, 
that is, at $\sim$40\% of the substrate melting temperature. At the same time,
the quenching rate is high (see Section \ref{experimental}). We believe that
in this case the overall outline of the ECS is preserved during quenching,
although reconfigurations within a few interatomic distances are possible.
We have verified that, for the crystal studied here, 
lowering of the quenching rate leads to a different shape:
the vertex is significantly sharper. 

Our view
that the overall outline of the ECS is preserved during quenching
may not be valid for temperatures above 1500~K, where surface diffusion
is very efficient. The images of O/W annealed at 1640--1800~K
show a complex structure \cite{SzczepkowiczBryl2004}, which may be due to
non-equilibrium processes occurring during quenching.

Although the thermal vertex rounding for adsorption systems has not been
reported previously, there appears to be a close correspondence of
the O/W vertex rounding described in this work,
and the results reported by Song et al.\ \cite{Song1995} for a planar O/Mo(111) system.
They have reported the reversible planar/faceted phase transition 
at $\sim950$~K (oxygen dosage: 0.8~L). A similar transition occurs in
Oleksy's Monte Carlo model \cite{Oleksy2004}. Oxygen-induced faceting is similar 
on Mo(111) and W(111) substrates \cite{Madey1990, Madey1993a, Song1995}, 
so a planar/faceted transition is also expected
for O/W(111) \cite{Song2004}. We believe that that the temperature of the planar/faceted
transition is equal to the vertex-rounding temperature of the ECS, if the adsorbate coverage
is the same. We hope that this hypothesis will be addressed both by theorists and 
experimentalists. 

In light of the observations of the planar/faceted transition, a question
arises: is the vertex \emph{rounded} for O/W ECS above 970 K, or is it
\emph{truncated} by a flat (111) facet?
It is not possible to address this question on the basis of our microscopic images,
because of insufficient microscope resolution, because of possible image
distortions \cite{FIM-Oxford, SzczepkowiczCiszewski2002}, and because of
the possibility of atomic-scale reconstruction of this area during quenching.
However, the Monte Carlo model \cite{Oleksy2004} shows that the phase 
above the transition temperature is not flat after all, exhibiting large variations
from the ideal BCC(111) structure, while still being able to produce the BCC(111) LEED 
spots (see Figs.~5(b) and 9 in \cite{Oleksy2004}). For this reason, we believe that
the thermal evolution of the oxygen-covered tungsten crystal can be effectively
described by the type-B scheme (rough, rounded vertex, sharp edges).

\begin{acknowledgments}
We would like to thank Dr.\ C. Oleksy, Dr.\ K.-J. Song and Prof.\ T.E. Madey
for stimulating discussions.
\end{acknowledgments}

\bibliography{vertex_rounding.bbl}

\end{document}